\documentstyle[12pt]{article}

\begin{document}
\setcounter{page}{1} \pagestyle{plain} \vspace{1cm}
\begin{center}
\Large{\bf Loop Quantum Gravity Modification of the Compton Effect}\\
\small
\vspace{1cm} {\bf Kourosh Nozari$^1$}\quad\ and \quad {\bf S. Davood Sadatian$^2$}\\
{$^{1,2}$\it Department of Physics,
Faculty of Basic Sciences,\\
University of Mazandaran,\\
P. O. Box 47416-1467, Babolsar, IRAN\\e-mail:\\
$^1$knozari@umz.ac.ir\\$^2$d.sadatian@umz.ac.ir }
\end{center}
\vspace{1.5cm}
\begin{abstract}
Modified dispersion relations(MDRs) as a manifestation of Lorentz
invariance violation, have been appeared in alternative approaches
to quantum gravity problem. Loop quantum gravity is one of these
approaches which evidently requires modification of dispersion
relations. These MDRs will affect the usual formulation of the
Compton effect. The purpose of this paper is to incorporate the
effects of loop quantum gravity MDRs on the formulation of Compton
scattering. Using limitations imposed on MDRs parameters from Ultra
High Energy Cosmic Rays(UHECR), we estimate the quantum
gravity-induced wavelength shift of scattered photons in a typical
Compton process. Possible experimental detection of this wavelength
shift will provide strong support for underlying quantum gravity proposal.
\end{abstract}
\newpage

\section{Introduction}
Historically, Compton effect is one of the most important evidence
of particle nature of electromagnetic radiation. Compton scattering
or equivalently Compton effect, is the reduction of energy ( or
increase of wavelength) of an X-ray or gamma ray photon, when it
interacts with matter. The amount the wavelength increases by is
called the Compton shift. Although nuclear Compton scattering
exists, what is meant by Compton scattering usually is the
interaction involving only the electrons of an atom. This effect is
important because it demonstrates that light cannot be explained
purely as a wave phenomenon. Thomson scattering, the classical
theory of charged particles scattered by an electromagnetic wave,
can not explain any shift in wavelength. Light must behave as if it
consists of particles in order to explain the Compton scattering.
Compton's experiment convinced physicists that light can behave as a
stream of particles whose energy are proportional to the
frequency[1]. When a high energy photon collides an electron, part
of its initial energy will transfer to electron and cause it to
recoil. The other part of initial photon energy leads to creation of
a new photon and this photon moves in a direction which satisfy the
total momentum conservation. Compton scattering occurs in all
materials and predominantly with photons of medium energy, i.e.
about 0.5 to 3.5 MeV.\\
Recently it has been revealed that Lorentz symmetry is not an exact
symmetry of the nature. Possible violation of Lorentz invariance has
been studied from several view points [2,3,4]. From a loop quantum
gravity point of view, a Lorentz invariance violation can be
formulated in the modification of standard dispersion relation.
Since formulation of ordinary Compton effect is based on the
standard dispersion relation, possible modification of this
dispersion relation may affect the calculations and their
interpretations. Here we are going to incorporate these quantum
gravity effects in the formulation of the Compton effect. Although
numerical values of these modifications are very small, possible
detection of these small effects will support underlying quantum
gravity proposal. We use UHECR data to estimate typical wave length
shift due to these quantum gravity effects. In this manner we
constraint threshold momentum to a lower value relative to existing
prescriptions.\\
The paper is organized as follows: section 2 is devoted to a brief
review of standard Compton effect and its formulation focusing on
the central role played by dispersion relation. Section 3 gives an
overview of modified dispersion relations. Our calculations are
presented in section 4. The paper follows by a numerical estimation
of the wavelength shift and related discussion.

\section{Compton Effect}
Suppose that a photon with known wavelength collides with a thin
metallic surface. In classical theory of light scattering, photon
will be reflected by oscillating electron in such a way that its
angular distribution varies as\, $1+cos^{2}\theta$. Compton had been
noticed that scattered photons consist of two different wavelength:
a part with the same wavelength as original photons. These photons
have been scattered by the whole of the atom. Another part of
scattered photons have shifted wavelength relative to incident
photons and their wavelength depend on the angular parameter. If we
consider an elastic collision between photon and electron, total
energy and momentum should be conserved in this process. Suppose
that incident photon has energy $h\nu$ and momentum $\vec{p}$ where
$p=\frac{h\nu}{c}$. From standard dispersion relation between energy
and momentum we have
\begin{equation}
E=[(m_0c^2)^2+(pc)^2]^\frac{1}{2},
\end{equation}
where $m_0$ is the rest mass of the particle. The speed of the
particle is given by
\begin{equation}
v=\frac{\partial E}{\partial
p}=\frac{pc^2}{E}=\frac{pc^2}{\sqrt{{m_0}^2c^4+p^2c^2}}
\end{equation}
where leads to \, $ E=pc$ \, for photons. Now consider a photon with
initial momentum $\vec{p}$ which collides with an electron at rest.
After collision, we have an electron with momentum $\vec{P}$ and a
new photon with momentum $\vec{p'}$. From momentum conservation one
can write
\begin{equation}
\vec{p}=\vec{p'}+\vec{P}
\end{equation}
which leads to
\begin{equation}
(\vec{P})^2=(\vec{p}-\vec{p'})^2=(\vec{p})^2+(\vec{p'})^2-2\vec{p}.\vec{p'}.
\end{equation}
From conservation of energy, one can write
\begin{equation}
h\nu-h\nu'=E-E_0=(P^2c^2+{E}^2_0)^\frac{1}{2}-E_0 ,
\end{equation}
where $E_0$ is electron rest energy. Therefore, we obtain
\begin{equation}
m^2c^4+P^2c^2=(h\nu-h\nu'+mc^2)^2=(h\nu-h\nu')^2+2mc^2(h\nu-h\nu')+m^2c^4
\end{equation}
Using equation (4) we can write
\begin{equation}
P^2=(\frac{h\nu}{c})^2+(\frac{h\nu'}{c})^2-2(\frac{h\nu}{c})(\frac{h\nu'}{c})cos\theta
\end{equation}
which means
\begin{equation}
P^2c^2=(h\nu-h\nu')^2+2(h\nu)(h\nu')(1-cos\theta)
\end{equation}
where $\theta$ is the angle of photon scattering. A simple
calculation leads to the following expression
\begin{equation}
h\nu\nu'(1-cos\theta)=mc^2(\nu-\nu')
\end{equation}
which can be rewritten as follows
\begin{equation}
\lambda'-{\lambda}=\frac{h}{mc}(1-cos\theta)
\end{equation}

\begin{figure}[htp]
\begin{center}
\includegraphics{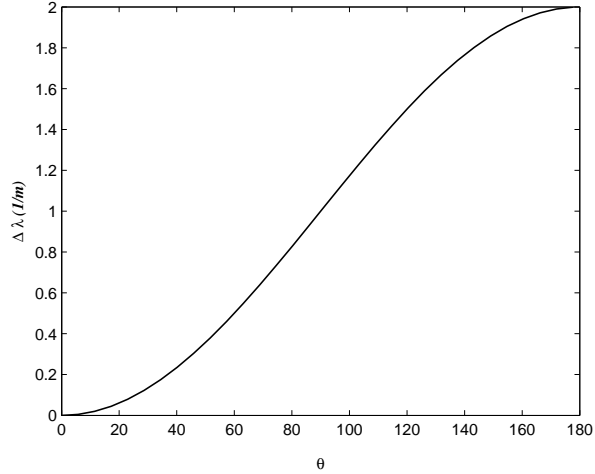}
\end{center}
\vspace{7cm} \caption{\small {Standard Compton Effect. }}
\label{fig:1}
\end{figure}
where \,$\frac{h}{mc}$\, is called the Compton wavelength. Note that
scattered electron and wavelength-shifted photon appear
simultaneously. Figure \,$1$ \,shows the variation of $\Delta
\lambda$ versus scattering angle $\theta$.
\\
\\
\\
\\
\\
\section{Modified Dispersion Relations (MDRs)}
Lorentz invariance violation at quantum gravity level can be
addressed by the modification of the standard dispersion relations.
Recently these modified dispersion relations have been used to
describe anomalies in astrophysical phenomena such as the GZK cutoff
anomaly[5,6] as well as a large number of problems in the spirit of
quantum mechanics(see for instance [5] and references therein).
Modified dispersion relations have support on several alternative
approaches to quantum gravity problem[7,8]. Here we consider those
MDRs formulations that contain Planck length ($l_{p}$) explicitly.
In this framework, there is a new length scale, $L\gg l_p$, which is
called "weave" scale. For distance $d\ll L$ the quantum loop
structure of spacetime is manifest, while for distances $d\gg L$ the
continuous flat geometry is regained[7]. In this context, dispersion
relation $E=E(p)$ for particle with energy $E$ and momentum $p$ is
given by ($\hbar=c=1$) [8,9]
\begin{equation}
E^2=A^2p^2+m^2
\end{equation}
where $E$, $p$ and $m$ are the respective energy, momentum and mass
of the particle and $A$ is a Lorentz invariance violation parameter
which can be interpreted as the maximum velocity of the
particle\footnote{It can be shown that these Lorentz invariance
violation can significantly modify the kinematical conditions for a
reaction to take place[8].}. For Fermions( Majorana Fermions), these
modified dispersion relations can be written as follows[7,8]
\begin{equation}
{E^2}_\pm=(Ap\pm\frac{B}{2L})^2+m^2(\alpha\pm\beta p)^2
\end{equation}
where
\begin{equation}
A=1+\kappa_1\frac{l_p}{L}+\kappa_2(\frac{l_p}{L})^2+\frac{\kappa_3}{2}{l_p}^2p^2
\end{equation}
\begin{equation}
B=\kappa_5\frac{l_p}{L}+\kappa_6(\frac
{l_p}{L})^2+\frac{\kappa_7}{2}{l_p}^2p^2
\end{equation}
\begin{equation}
\alpha=(1+\kappa_8\frac{l_p}{L})~~~~~~ and~~~~~~
\beta=\frac{\kappa_9}{2}l_p
\end{equation}
Here $\kappa_i$ are unknown adimensional parameters of order one and
the\, $\pm$\, signs stand for the helicity of propagation. For
simplicity we write
\begin{equation}
{E_\pm}^2=A^2p^2+\eta p^4\pm 2\Gamma p+m^2
\end{equation}
where now $A=1+\frac{\kappa_1 l_p}{L}$ \,and \,$\kappa_1$,\,
$\kappa_3$ and  $\kappa_5$ are of the order of unity with
$\eta=\kappa_3{l_p}^2$ and $\Gamma=\kappa_5 \frac{l_p}{2L^2}$ which
depends on the helicity.\\
The following MDR has been suggested also
\begin{equation}
E^2=p^2+m^2+\frac{|p|^{2+n}}{M^n}
\end{equation}
where $M^{n}$ is the characteristic scale of Lorentz
violation[9,10]. In which follows, we use these relations to
incorporate quantum gravitational effects in the formulation of the
Compton effect. The possible values of parameters in equation (16)
have been discussed in references[7,8] using UHECR data. We use
these values to estimate the order of wavelength shift due to loop
quantum gravity effect.

\section{Generalized Compton Effect}
In this section we use modified dispersion relations as given by
(16) and (17) to incorporate quantum gravitational effects in the
formulation of the Compton scattering. In a typical Compton
scattering, the conservation of linear momentum leads to
\begin{equation}
\vec{p}=\vec{p'}+\vec{P}
\end{equation}
where \,$\vec{p}$\, is linear momentum of the incident photon, \,
$\vec{p'}$ is momentum of the wavelength-shifted secondary photon
and $\vec{P}$ is electron linear momentum. In this situation, we can
write
\begin{equation}
\vec{P^2}=(\vec
{p}-\vec{p'})^2=\vec{p^2}+\vec{p'^2}-2\vec{p}.\vec{p'}
\end{equation}
In the first step, we assume that photon dispersion relation has no
loop quantum gravity modification. This assumption will be justified
later. For conservation of energy, one should consider modified
dispersion relation for electron. If we use relation (16), we find
\begin{equation}
\nu-\nu'=E-E_0=(A^2P^2+\eta P^4\pm 2\Gamma P+m^2)^{\frac{1}{2}}-m
\end{equation}
where we have set $\hbar=c=1$. \, $m=E_0$ is rest energy and $E$ is
the final energy of the electron. If we rearrange this relation we
find
\begin{equation}
(\nu-\nu'+m)^2=A^2P^2+\eta P^4\pm2\Gamma P+m^2
\end{equation}
where leads to the following expression
\begin{equation}
P^2=\frac{1}{A^2}\Big[(\nu-\nu')^2+2(\nu-\nu')m-\eta P^4\mp \Gamma
P\Big]
\end{equation}
We can write \, $A=1+\epsilon$ \, where \, $\epsilon =
\frac{\kappa_1 l_p}{L}$\, is a small quantity, then using the
formula \, $(1+\epsilon)^{-2}\simeq 1-2\epsilon $\, we find
\begin{equation}
P^2=(\nu-\nu')^2+2(\nu-\nu')m-\eta P^4\mp \Gamma
P-2\epsilon(\nu-\nu')^2-4\epsilon(\nu-\nu')m+2\epsilon\eta P^4\pm2\epsilon\Gamma
P
\end{equation}
From equation (19) we can write
\begin{equation}
P^2=(\nu-\nu')^2+2\nu\nu'(1-cos\theta)
\end{equation}
Finally, combining equations (23) and (24) we find
\begin{equation}
2\nu\nu'(1-cos\theta)=(1-2\epsilon)\Big[2(\nu-\nu')m-\eta
P^4\mp\Gamma P\Big]-2\epsilon(\nu-\nu')^2
\end{equation}
The frequency shift of the scattered photon can be obtained using
this relation. If we look at equation (25) and demand that both
photon frequencies be positive, then we learn that $\eta$ should be
negative and one of the polarizations is ruled out. This is in
agreement with the result of ref.[7]. Equation (25) contains some
parameters which should be constraint using phenomenological
evidences of these
quantum gravity effects.\\
We can proceed also using relation (17)
for modified dispersion relation. A simple calculation leads to the
following result
\begin{equation}
P^2=(\nu-\nu')^2+2m(\nu-\nu')-\frac{|P|^{2+n}}{M^n}
\end{equation}
which can be transformed to the relation
\begin{equation}
2\nu\nu'(1-cos\theta)=2m(\nu-\nu')-\frac{|P|^{2+n}}{M^n}
\end{equation}
The quantum gravity-corrected Compton effect now takes the following
form
\begin{equation}
\lambda'-\lambda=\frac{1}{m}(1-cos\theta)+\frac{\lambda\lambda'
|P|^{2+n}}{2M^nm}
\end{equation}
where $M^{n}$  is the characteristic scale of Lorentz invariance
violation. The matter which should be stressed is the fact that
wavelength shift due to loop quantum gravity effect is wavelength
dependent itself. In another words, the value of this shift depends
on the wave length of incident photon. This is a novel feature. In
ordinary
Compton effect such a wavelength dependence has not been observed.\\
Relation (28) can be written as follows
\begin{equation}
\lambda'=\frac{\lambda+\frac{1}{m}(1-cos\theta)}{1-\frac{\lambda|P|^{2+n}}{2M^nm}}.
\end{equation}
Note that equation (29) should be supplemented by an arbitrary
factor $F$ of order $1$, then equation (26) will say that $F<0$ and
the contradiction of having negative wavelength is avoided.

In $MKS$ system of units, this relation takes the following form
\begin{equation}
\lambda'=\frac{\lambda+\frac{h}{mc}(1-cos\theta)}{1-\frac{\lambda|Pc|^{2+n}}{2h(Mc^2)^{n}mc^3}}.
\end{equation}
Since for reasonable values of $|P|$, \, the
quantity\,\,$\frac{\lambda|Pc|^{2+n}}{2h(Mc^2)^{n}mc^3}$ \, is a
positive small quantity, loop quantum gravity induces an increase of
wavelength shift. Figure $2$ shows the variation of $\lambda'$
versus $\theta$ based on equation (30).\\

Now we consider the more general case where photon dispersion
relation is modified by quantum gravity effect also. For photons,
the modified dispersion relation can be written as [11]
\begin{equation}
E_{\pm}=p[A_\gamma-\theta_3(l_pp)^2\pm\theta_8l_pp]
\end{equation}
where
\begin{equation}
A_\gamma=1+\kappa_\gamma(\frac{l_p}{L})^{2+2\Upsilon}
\end{equation}
In this relation, $E_{\pm}$ and $p$ are the respective energy and
momentum of the photon, while $\kappa_{\gamma}$ and $\theta_{i}$ are
adimensionl parameters of order one, and $\Upsilon$ is a free
parameter that still needs interpretation\footnote{ It should be
noted that the presence of the $\Upsilon$ parameter in the fermion
dispersion relation was not considered in some literature such as
[12]}. A similar contribution was also suggested by Ellis {\it et
al} [13],[14] (in this case, without helicity dependance). They have
found
\begin{equation}
E^{2} = p^{2} \left[ 1- 2 M_{D}^{-1} p \right],
\end{equation}
where $M_{D}$ is a mass scale coming from D-brane recoil effects for
the propagation of photons in vacuum. When Gamma Ray Burst (GRB)
data are analyzed to restrict $M_{D}$ [14], the following condition
arises
\begin{equation}
{M_{D}}\geq {10^{24}}  eV
\end{equation}
For photon's dispersion relation which we have considered, (32) can
be interpreted as the bound ${\theta_{\gamma}} \leq{10^{4}}$. To
consider photon(Boson) dispersion relation in our analysis, we start
with equation (20) and we obtain
\begin{equation}
p[A_\gamma-\theta_3(l_pp)^2\pm\theta_8l_pp]-p'[A_\gamma-\theta_3(l_pp')^2\pm\theta_8l_pp']=E-E_0=(A^2P^2+\eta
P^4\pm 2\Gamma P+m^2)^{\frac{1}{2}}-m
\end{equation}
This equation has several parameters which should be constraint from
experimental or observational data. Another form of this relation
can be obtain using equation (31)
\begin{equation}
p \left[ 1- 2 M_{D}^{-1} p \right]^\frac{1}{2}-p' \left[ 1- 2
M_{D}^{-1} p' \right]^\frac{1}{2}=E-E_0=(A^2P^2+\eta P^4\pm 2\Gamma
P+m^2)^{\frac{1}{2}}-m.
\end{equation}
These two relations are more general than (25) and (29). One can use
this relations for an estimation of induced wavelength shift.
However, in which follows, for simplicity we consider only relation
(30) to estimate a typical wavelength shift .
\section{Threshold Analysis}
In this paper we have focused on an interaction which contains
photons and electrons. In this way, we are able to obtain rather
strong constraints on the allowed parameter space from threshold
analysis[18]. Based on these analysis the photon decay rate goes
like $E$ above threshold, so any gamma ray which propagates over
macroscopic distances must have energy below the threshold. This
threshold now is supposed to be $|{p}_{th}|= 10^{13} eV $ [17,18].
If we accept this threshold, we can estimate the value of the
wavelength shift due to loop quantum gravity effect in a typical
Compton effect. For simplicity we consider relation (30). In this
relation, $M$ causes the appearance of threshold effects at momenta
$|{p}| \geq |{p}_ {th}|$, where $|{p}_{th}|^{2+n} = m^2 M^n$. It is
these effects which allows to explore quantum gravity at energies
much lower than $M$ [15]. Carmona and  Cortes have argued that for
$M=M_P$ and a typical hadronic process, one gets $|{p}_{th}|=
10^{15} eV$ in the case $n=1$ and $|{p}_{th}|= 10^{18} eV$ in the
case $n=2$. In both cases, one has modifications to relativistic
kinematics at energies below the GZK cutoff, so that the observed
violations of this cutoff in the cosmic ray spectrum[15] could be a
footprint of a Lorentz invariance violation at high energies[8].
First we try to find a wavelength shift using Carmona-Cortes
threshold. Suppose that a photon with wavelength
$0.71{\AA}=0.71\times10^{-10}m$(which is the wavelength of photon in
original Compton experiment) is scattered by electron via Compton
process at angle $\theta=\frac{\pi}{2}$. Using the relation (30) we
find an unacceptable wavelength shift: an extremely small negative
shift which is unacceptable due to its negative sign. This may
reflect the fact that Carmona-Cortes threshold is not acceptable on
physical ground. Using relation (30) we can obtain a reasonable
wavelength shift of the scattered photon. Our calculations show a
shifted wavelength of $\lambda'=7.376636\times 10^{-11} m$ which
leads to $\Delta \lambda_{LQG}= 0.033636\times 10^{-11}m$. Note that
we have used $|P|\leq 10^{12} eV$ to have a reasonable wavelength
shift. Looking back to relation (30) we see that the only arbitrary
quantity is $|P|$. In our situation, only when $|P|\leq 10^{12} eV$
we find a reasonable wavelength shift. This threshold is one order
of magnitude smaller than threshold presented by Jacobson {\it et
al}. So, we may conclude that the reasonable threshold should be
$|P|\leq 10^{12} eV$. Note that $0.343\times 10^{-11} m$ is the
ordinary wavelength shift via Compton process. In summary, we
conclude that to have a reasonable wavelength shift due to loop
quantum gravity effect, threshold effects should take place at
$|{p}_{th}|= 10^{12} eV$. For comparison, note that Carmona and
Cortes[10] have considered $|{p}_{th}|= 10^{15} eV$ for $n=1$ while
Jacobson {\it et al} have considered  $|{p}_{th}|= 10^{13} eV$[17].
Our analysis shows that Jacobson {\it et al} framework gives more
reliable result of threshold effect. Finally, we should emphasize
that Lowering of this threshold may affect several of arguments in
this research area.
\\
\\
\\

\begin{figure}[htp]
\begin{center}
\includegraphics{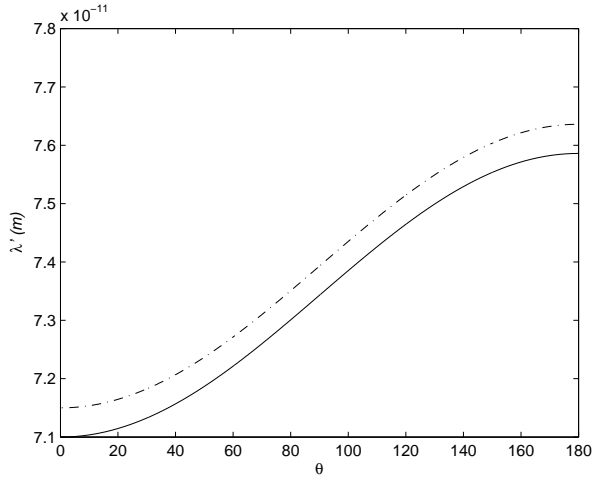}
\end{center}
\vspace{5cm} \caption{\small {Variation of $\lambda'$ in modified
Compton effect for X-rays with wavelength $0.71{\AA}$. Due to
smallness of the modification, it has been multiplied with an
arbitrary factor.}} \label{fig:2}
\end{figure}

\section{summary}
The results of our analysis of Compton effect within the framework
of modified dispersion relations can be summarized as follows:
\begin{itemize}
\item
The wavelength shift due to loop quantum gravity effect is
wavelength dependent itself. In another words, the value of the
wavelength shift depends on the wavelength of incident photon.
\item
Through a threshold analysis we have found a threshold for the
momentum of electron as $|P|\leq 10^{12} eV$. Our analysis shows
that Jacobson {\it et al} framework[17,18] gives more reliable
result of threshold effect.
\item
The analysis presented here provides a direct test of loop quantum
gravity and Lorentz invariance violation. Any wavelength shift of
scattered photon after subtraction of ordinary Compton shift will
show the violation of Lorentz invariance and provides a direct test
of loop quantum gravity.
\end{itemize}

\end{document}